\documentclass[3p,twocolumn]{elsarticle}
\usepackage{graphicx}
\usepackage{epsfig}
\usepackage{amsmath}
\usepackage{dcolumn}
\usepackage{bm}
\usepackage{dcolumn}
\usepackage{color}
\usepackage{verbatim}
\usepackage{paralist}
\usepackage{amssymb}
\usepackage{url}
\begin{document}
\newcommand{\exposure}{252.6\,ton$\cdot$yr}
\newcommand{\unit}{events/(100\,ton$\cdot$yr)}
\begin{frontmatter}
\title{Measurement of geo--neutrinos from 1353 days of Borexino} 
\author[Milano]{G.~Bellini}
\author[PrincetonChemEng]{J.~Benziger}
\author[Hamburg]{D.~Bick}
\author[LNGS]{G.~Bonfini}
\author[Virginia]{D.~Bravo}
\author[APC]{M.~Buizza~Avanzini}
\author[Milano]{B.~Caccianiga}
\author[UMass]{L.~Cadonati}
\author[Princeton]{F.~Calaprice}
\author[LNGS]{P.~Cavalcante}
\author[Princeton]{A.~Chavarria}
\author[Lomonosov]{A.~Chepurnov}
\author[Milano]{D.~D'Angelo}
\author[Houston]{S.~Davini}
\author[Peters]{A.~Derbin}
\author[Houston]{A.~Empl}
\author[Kurchatov]{A.~Etenko}
\author[Ferrara]{G.~Fiorentini}
\author[Dubna]{K.~Fomenko}
\author[APC]{D.~Franco}
\author[Princeton]{C.~Galbiati}
\author[LNGS]{S.~Gazzana}
\author[APC]{C.~Ghiano}
\author[Milano]{M.~Giammarchi}
\author[Munich]{M.~Goeger-Neff}
\author[Princeton]{A.~Goretti}
\author[Princeton]{L.~Grandi}
\author[Hamburg]{C. Hagner}
\author[Houston]{E.~Hungerford}
\author[LNGS]{Aldo~Ianni}
\author[Princeton]{Andrea~Ianni}
\author[Kiev]{V.V.~Kobychev}
\author[Dubna]{D.~Korablev}
\author[Houston]{G.~Korga}
\author[LNGS]{Y.~Koshio}
\author[APC]{D.~Kryn}
\author[LNGS]{M.~Laubenstein}
\author[Munich]{T.~Lewke}
\author[Kurchatov]{E.~Litvinovich}
\author[Princeton]{B.~Loer}
\author[Milano]{P.~Lombardi}
\author[LNGS]{F.~Lombardi}
\author[Milano]{L.~Ludhova}
\author[Kurchatov]{G. Lukyanchenko}
\author[Kurchatov]{I.~Machulin}
\author[Virginia]{S.~Manecki}
\author[Heidelberg]{W.~Maneschg}
\author[Ferrara]{F.~Mantovani}
\author[Genova]{G.~Manuzio}
\author[Munich]{Q.~Meindl}
\author[Milano]{E.~Meroni}
\author[Milano]{L.~Miramonti}
\author[Krakow]{M.~Misiaszek}
\author[Princeton]{P.~Mosteiro}
\author[Peters]{V.~Muratova}
\author[Munich]{L.~Oberauer}
\author[APC]{M.~Obolensky}
\author[Perugia]{F.~Ortica}
\author[UMass]{K.~Otis}
\author[Genova]{M.~Pallavicini}
\author[Virginia]{L.~Papp}
\author[Genova]{L.~Perasso}
\author[Genova]{S.~Perasso}
\author[UMass]{A.~Pocar}
\author[Milano]{G.~Ranucci}
\author[LNGS]{A.~Razeto}
\author[Milano]{A.~Re}
\author[Ferrara]{B.~Ricci}
\author[Perugia]{A.~Romani}
\author[LNGS]{N.~Rossi}
\author[Kurchatov]{A.~Sabelnikov}
\author[Princeton]{R.~Saldanha}
\author[Genova]{C.~Salvo}
\author[Munich]{S.~Sch\"onert}
\author[Heidelberg]{H.~Simgen}
\author[Kurchatov]{M.~Skorokhvatov}
\author[Dubna]{O.~Smirnov}
\author[Dubna]{A.~Sotnikov}
\author[Kurchatov]{S.~Sukhotin}
\author[UCLA,Kurchatov]{Y.~Suvorov}
\author[LNGS]{R.~Tartaglia}
\author[Genova]{G.~Testera}
\author[APC]{D.~Vignaud}
\author[Virginia]{R.B.~Vogelaar}
\author[Munich]{F.~von~Feilitzsch}
\author[Munich]{J.~Winter}
\author[Krakow]{M.~Wojcik}
\author[Princeton]{A. Wright}
\author[Hamburg]{M.~Wurm}
\author[Princeton]{J.~Xu}
\author[Dubna]{O.~Zaimidoroga}
\author[Genova]{S.~Zavatarelli}
\author[Krakow]{G.~Zuzel}
\author{\\(Borexino Collaboration)}

\address[Milano]{Dipartimento di Fisica, Universit\`a degli Studi and INFN, 20133 Milano, Italy}
\address[PrincetonChemEng]{Chemical Engineering Department, Princeton University, Princeton, NJ 08544, USA}
\address[Hamburg]{University of Hamburg, 22761 Hamburg, Germany}
\address[LNGS]{INFN Laboratori Nazionali del Gran Sasso, SS 17 bis Km 18+910, 67010 Assergi (AQ), Italy}
\address[Virginia]{Physics Department, Virginia Polytechnic Institute and State University, Blacksburg, VA 24061, USA}
\address[APC]{APC, Universit\' e Paris Diderot, CNRS/IN2P3, CEA/Irfu, Obs de Paris, Sorbonne Paris Cit\' e, France}
\address[UMass]{Physics Department, University of Massachusetts, Amherst, MA 01003, USA}
\address[Princeton]{Physics Department, Princeton University, Princeton, NJ 08544, USA}
\address[Lomonosov]{Lomonosov Moscow State University, Skobeltsyn Institute of Nuclear Physics, Moscow 119234, Russia}
\address[Houston]{Department of Physics, University of Houston, Houston, TX 77204, USA}
\address[Peters]{St. Petersburg Nuclear Physics Institute, 188350 Gatchina, Russia}
\address[Kurchatov]{NRC Kurchatov Institute, 123182 Moscow, Russia}
\address[Ferrara]{Dipartimento di Fisica e Scienze della Terra, Universit\`a degli Studi and INFN, Ferrara I-44122, Italy}
\address[Dubna]{Joint Institute for Nuclear Research, 141980 Dubna, Russia}
\address[Munich]{Physik Department, Technische Universit\"at Muenchen, 85748 Garching, Germany}
\address[Kiev]{Institute for Nuclear Research, 03680 Kiev, Ukraine}
\address[Heidelberg]{Max-Planck-Institut f\"ur Kernphysik, 69029 Heidelberg, Germany}
\address[Genova]{Dipartimento di Fisica, Universit\`a and INFN, Genova 16146, Italy}
\address[Krakow]{M.~Smoluchowski Institute of Physics, Jagiellonian University, 30059 Cracow, Poland}
\address[Perugia]{Dipartimento di Chimica, Universit\`a e INFN, 06123 Perugia, Italy}
\address[UCLA]{Physics and Astronomy Department, University of California Los Angeles, Los Angeles, CA 90095, USA}

\begin{abstract}
We present a measurement of the geo--neutrino signal obtained from 1353 days of data with the Borexino detector at Laboratori Nazionali del Gran Sasso in Italy.  With a fiducial exposure of (3.69 $\pm$ 0.16) $\times$ $10^{31}$ proton $\times$ year after all selection cuts and background subtraction, we detected (14.3 $\pm$ 4.4) geo--neutrino events assuming a fixed chondritic mass Th/U ratio of 3.9. This corresponds to a geo--neutrino signal $S_{geo}$ = (38.8 $\pm$ 12.0)~TNU with just a 6 $\times$ $10^{-6}$ probability for a null geo--neutrino measurement. With U and Th left as free parameters in the fit, the relative signals are $S_{\mathrm{Th}}$ = (10.6 $\pm$ 12.7)~TNU and $S_\mathrm{U}$ = (26.5 $\pm$ 19.5)~TNU. Borexino data alone are compatible with a mantle geo--neutrino signal of (15.4 $\pm$ 12.3)~TNU, while a combined analysis with the KamLAND data allows to extract a mantle signal of (14.1 $\pm$ 8.1)~TNU. Our measurement of 31.2$^{+7.0}_{-6.1}$ reactor anti--neutrino events is in agreement with expectations in the presence of neutrino oscillations.
\end{abstract}

\end{frontmatter}
Geo--neutrinos (geo--$\bar{\nu}_{e}$) are electron anti--neutrinos ($\bar{\nu}_{e}$) produced mainly in $\beta$ decays of $^{40}$K and of several nuclides in the chains of long--lived radioactive isotopes $^{238}$U and $^{232}$Th, which are naturally present in the Earth.  
By measuring the geo--neutrino flux from all these elements, it is in principle possible to deduce the amount of the radiogenic heat produced within the Earth, an information facing large uncertainty and being of crucial importance for geophysical and geochemical models.
The first experimental investigation of geo--neutrinos from $^{238}$U and $^{232}$Th was performed by the KamLAND collaboration~\cite{FirstGeoKam1, FirstGeoKam2}, followed by their observation with a high statistical significance of 99.997\% C.L. by Borexino~\cite{BorexGeonu} and KamLAND\footnote{An update of geoneutrino analysis from KamLAND collaboration~\cite{KamNew} was released after the submission of this paper and thus it was not considered in the combined analysis presented below.}~\cite{gando}.
Both these experiments are using large--volume liquid--scintillator detectors placed in underground laboratories shielded against cosmic muons.
Due to either low statistics and/or systematic errors, these measurements do not have the power to discriminate among several geological models.
Analysis combining the results from different sites have higher prediction power, as it was shown in~\cite{gando} and \cite{Fiorentini2012}. 
Therefore, new measurements of the geo--neutrino flux are highly awaited by this newly--born inter--disciplinary community.
Several projects entering operation such as SNO+~\cite{SNO+} or under the design phase as LENA~\cite{LENA} or Hanohano~\cite{HanoHano}, have geo--neutrinos among their scientific aims.
In this work we present a new Borexino measurement of the geo--neutrino signal with 2.4 times higher exposure with respect to~\cite{BorexGeonu}. 
For the first time, Borexino attempts a measurement of the individual geo--neutrino signals from the $^{238}$U and $^{232}$Th chains.
We provide a detailed comparison of our measurement with the predictions of several geological models.
In a combined analysis of the Borexino and KamLAND~\cite{gando} data we provide an estimate of the mantle geo--neutrino signal.

Borexino is an unsegmented liquid scintillator detector built for the spectral measurement of low--energy solar neutrinos installed in the underground hall C of the Laboratori Nazionali del Gran Sasso (LNGS) in Italy.  
Several calibration campaigns with radioactive sources~\cite{BorexCalib} allowed us to decrease the systematic errors of our measurements and to optimize the values of several input parameters of the Monte--Carlo (MC) simulation.
The 278\,tons of ultra--pure liquid scintillator (pseudocumene (PC) doped with 1.5~g/l of diphenyloxazole) are confined within a thin spherical nylon vessel with a radius of 4.25\,m. 
The detector core is shielded from external radiation by 890 tons of buffer liquid, a solution of PC and 3-5~g/l of the light quencher dimethylphthalate.
The buffer is divided in two volumes by the second nylon vessel with a 5.75~m radius, preventing inward radon diffusion.
All this is contained in a 13.7~m diameter stainless steel sphere (SSS) on which are mounted 2212~8''~PMTs detecting the scintillation light, the so--called Inner Detector.  
An external domed water tank of 9\,m radius and 16.9\,m height,
filled with ultra--high purity water, serves as a passive shield
against neutrons and gamma rays as well as an active muon veto.
The Cherenkov light radiated by muons passing through the water is measured by 208~8" external PMTs also mounted on the SSS and define the so called Outer Detector.
A detailed description of the Borexino detector can be found in~\cite{BXdetector1,BXdetector2}.


In liquid scintillator detectors, $\bar{\nu}_{e}$ are detected via the inverse neutron $\beta$ decay,
\begin{equation}
\bar{\nu}_e + p \rightarrow e^+ + n,
\label{Eq:InvBeta}
\end{equation}
with a threshold of 1.806\,MeV, above which lies only a small fraction of $\bar{\nu}_{e}$ from the $^{238}$U (6.3\%) and $^{232}$Th (3.8\%) series.
Geo--neutrinos emitted in $^{40}$K decay cannot be detected by this technique.
The positron created in this reaction promptly comes to rest and annihilates.
All deposited energy is detected in a single prompt event, with a visible energy of $E_{\rm prompt} = E_{\bar{\nu}_e} - 0.784\,{\rm MeV}$. 
The emitted free neutron is typically captured on protons, resulting in the emission of a 2.22\,MeV de--excitation $\gamma$ ray, providing a delayed coincidence event.
The mean neutron capture time in Borexino was measured with an AmBe neutron source to be $\tau$ = (254.5 $\pm$ 1.8)\,$\mu$s~\cite{BorexMuons}.
The characteristic time and spatial coincidence of prompt and delayed events offers a clean signature of $\bar{\nu}_e$ detection, further suppressing possible background sources.

In this paper we report the analysis of data collected between December 2007 and August~2012, corresponding to 1352.60\,days of live time.  The fiducial exposure after all cuts is (613 $\pm$ 26) ton $\times$ year or (3.69 $\pm$ 0.16) $\times$ $10^{31}$ proton $\times$ year.

The $\bar{\nu}_e$'s from nuclear power plants are the main anti--neutrino background to the geo--neutrino measurement.
Since there are no nuclear plants close--by, the LNGS site is well suited for geo--neutrino detection.
The expected number of events from reactor $\bar{\nu}_e$'s, $N_{\rm react}$, is given by:
\begin{equation}
\begin{split}
N_{react}= & \sum_{r=1}^{R} \sum_{m=1}^{M} 
\frac {\eta_{m}}{4 \pi L_{r}^{2}} P_{rm}  \times   \\
\times &  \int dE_{\bar{\nu}_e} \sum_{i=1}^4 \frac {f_{i}}{E_{i}} \phi_{i}(E_{\bar{\nu}_e}) 
 \sigma(E_{\bar{\nu}_e})
P_{ee}(E_{\bar{\nu}_e};\hat\theta, L_r),
\end{split}
\label{Eq:ReactorFlux}
\end{equation}
where the index $r$ runs over the number $R$ of reactors considered, the index $m$ runs over the total number of months $M$ for the present data set, $\eta_m$ is the exposure (in proton $\times$ yr) in the $m^{\rm th}$ month including detector efficiency, $L_{r}$ is the distance of the detector from reactor $r$, $P_{rm}$ is the effective thermal power of reactor 
$r$ in month $m$, the index $i$ stands for the i-th spectral component in the set 
($^{235}$U, $^{238}$U, $^{239}$Pu, and $^{241}$Pu), $f_{i}$ is the power fraction of the component $i$,
$E_i$ is the average energy released per fission of the component $i$, 
$\phi(E_{\bar{\nu}})$ is the anti-neutrino spectrum per fission of the $i^{\rm th}$ component, 
$\sigma(E_{\bar{\nu}})$ is the inverse $\beta$ decay cross section taken from~\cite{Strumia}, 
and $P_{ee}$ is the survival probability~\cite{Fiorentini2012} of the reactor anti-neutrinos of energy 
$E_{\bar{\nu}}$ traveling the baseline $L_r$, for mixing parameters 
$\hat\theta$ = ($\delta m^2$, $\sin^2\theta_{12}$, $\sin^2\theta_{13}$).

In Eq.~(\ref{Eq:ReactorFlux}) we consider the $R$ = 446 nuclear cores all over the world,
operating in the period of interest. 
The mean weighted distance of these reactors from the LNGS site is about 1200\,km, being the weight $w_{rm}$ = $P_{rm} / L_r^2$.
The effective thermal power, $P_{rm}$, was calculated as a product of the nominal thermal power and the monthly load factor provided for each nuclear core by the International Atomic Energy Agency (IAEA)~\cite{IAEA}.
For each core the distance, $L_r$, has been calculated taking into account the geographic coordinates of the center of the Borexino detector (42.4540$^\circ$ latitude and 13.5755$^\circ$ longitude), obtained during the geodesy campaign for a measurement of CNGS muon--neutrino speed~\cite{BorexinoCNGS}.
The $\phi_i(E_{\bar{\nu}_e})$ energy spectra are taken from~\cite{Mueller2011}, differing  from the spectra~\cite{Huber2004} used in~\cite{BorexGeonu} by about +3.5\% in the normalization.
The shapes are comparable in the energy window of our anti--neutrino candidates. 
Note that the 3.5\% difference in the normalization is conservatively considered as a systematic error.
For the power fractions, $f_{i}$, we adopt the same assumptions as in our previous study~\cite{BorexGeonu}.
Furthermore, in this analysis we precise $f_{i}$ for the 46 cores using heavy water moderator~\cite{Candubook}:
\begin{equation}
\begin{split}
\mbox{$^{235}$U} & :\mbox{$^{238}$U}:\mbox{$^{239}$Pu}:\mbox{$^{241}$Pu}  = \\
0.542 & :0.411:0.022:0.0243  \\
\end{split}
\end{equation}
Since only two such cores are in Europe (in Romania) this improvement in the calculation has an effect less than 0.1\%.

We adopt neutrino oscillations parameters as derived in~\cite{Fogli2012} for normal hierarchy:
$\delta m^2$ = (7.54 $^{+0.26}_{-0.22}$)$\cdot 10^{-5}$ eV$^2$; 
$\sin^2\theta_{12}$ = (3.07 $^{+0.18}_{-0.16}$)$\cdot 10^{-1}$;
$\sin^2\theta_{13}$ = (2.41 $\pm$ 0.25)$\cdot 10^{-2}$.
The three flavor scenario implies a 4.6\% decrease in the predicted signal with respect the two neutrino case (as it was used in~\cite{BorexGeonu}), while the spectral shape does not significantly change.

As in~\cite{BorexGeonu}, we also include a +0.6\% contribution from matter effects (oscillation parameters as above), and the +1.0\% contribution of long--lived fission products in the spent fuel~\cite{Kopeikin}.  
The contributions to the estimated systematic error are summarized in Table~\ref{Tab:ReactorError}.

Finally, the number of expected reactor $\bar{\nu}_e$ candidates is $N_{react}$ = (33.3 $\pm$ 2.4) events for the exposure of (613 $\pm$ 26) ton $\times$ yr after cuts (for their efficiency see below).
We note that in the absence of oscillation, the number of expected events would be $60.4 \pm 4.1$. 

\begin{table}[h]
\caption{Systematic uncertainties on the  expected reactor--$\bar{\nu}_e$ signal which are added in quadrature.  See Eq.~(\ref{Eq:ReactorFlux}) and accompanying text for details.}
\begin{center}
\begin{tabular}{lr}
\hline\hline
Source								&Uncertainty	 \\
									&[\%]         \\ 
\hline
$\phi(E_{\bar{\nu}})$				        &3.5         \\
Fuel composition						&3.2		\\
$\theta_{12}$				                        & 2.3  \\
$P_{rm}$				                              &2.0\\
Long--lived isotopes						&1.0		\\
$E_i$					                      & 0.6\\
$\theta_{13}$				                       & 0.5\\
$L_{r}$					                     &0.4 \\
$\sigma_{\bar{\nu}p}$					&0.4		\\
$\delta m^2$						&0.03	\\
\hline
Total					      		&5.8 \\
\hline\hline
\end{tabular}
\label{Tab:ReactorError} 
\end{center}
\end{table}

The Borexino calibration campaigns~\cite{BorexCalib} included several $\gamma$, $\beta$, and $\alpha$ sources placed through the scintillator volume on and off-axis. The AmBe source, producing $\sim$10~neutrons/s with energies up to 10\,MeV, was deployed in twenty-five different positions allowing the study of the detector response to captured neutrons and to protons recoiling off neutrons. The calibration data were essential for testing the accuracy of the {\tt Geant4}--based Borexino MC simulation. The energy spectra of geo--neutrinos from  $^{238}$U and $^{232}$Th, based on the theoretical energy spectra of $\beta^{-}$ decays and the calculated energy spectrum of reactor $\bar{\nu}_e$'s (see above), were used as input to the MC code in order to simulate the detector response to $\bar{\nu}_e$ interactions. The MC output functions expressed in the total light yield, $Q$, (in units of photoelectrons, p.e., collected by the PMTs where 1 MeV corresponds to about 500 p.e.) were then used as fit functions in the final analysis. 
In this way, the non--linearities of the detector response function important at higher energies and in the increased fiducial volume with respect to solar neutrino analysis, are automatically taken into account.


The following cuts are used to select $\bar{\nu}_e$'s candidates: 1) $Q_{\rm prompt}$ $>$ 408\,p.e. and 860\,p.e. $<$ $Q_{\rm delayed}$ $<$ 1300~p.e., where Q$_{\rm prompt}$ and Q$_{\rm delayed}$ are the PMTs' light yields for the prompt (positron candidate) and delayed (neutron candidate) events; 2) reconstructed distance $\Delta R$ $<$ 1\,m; and 3) time interval 20\,$\mu$s $<$ $\Delta t$ $<$ 1280\,$\mu$s between the prompt and the delayed event. 
In liquid scintillators, a pulse--shape analysis can be used to discriminate highly ionizing particles ($\alpha$, proton) from particles with lower specific ionization ($\beta^-$, $\beta^+$, $\gamma$).
The so--called Gatti parameter $G$~\cite{gatti} has been used to improve background rejection.
For the delayed candidate a very slight cut requiring $G_{\rm delayed}$ $<$ 0.015 is applied.
The total detection efficiency with these cuts was determined by MC to be 0.84 $\pm$ 0.01. 

A minimal distance of 25~cm from the inner vessel containing the scintillator is required for the position of the prompt candidate.
Since this vessel is not perfectly spherical and does change in time, a dedicated algorithm was developed to calculate the vessel shape based on the position reconstruction of the events from the vessel's radioactive contaminants.
Since the vessel contamination is low, the vessel shape can be calculated only on a weekly basis. The precision of this method is 1.6\%. It was calibrated by comparing the vessel shapes with those obtained by a dedicated LED calibration system~\cite{BorexCalib}. The systematic error on the position reconstruction of $\bar{\nu}_e$ candidates is 3.8\%~\cite{BorexGeonu}. The total exposure of (613 $\pm$ 26) ton $\times$ year is calculated as a sum of weekly exposures which consider the corresponding weekly live time and the vessel shape as well as the (0.84 $\pm$ 0.01) efficiency of the selection cuts described above. The 4.2\% error on the exposure is a sum in quadrature of the errors on the vessel shape (1.6\%), on the position reconstruction of the candidates (3.8\%), and on the cuts efficiency (1\%).

Backgrounds faking anti--neutrino interactions can arise from cosmic muons and muon--induced unstable nuclides, from intrinsic contaminations of the scintillator and of the surrounding materials, and from the accidental coincidences of non-correlated events. A complete list of all expected backgrounds is reported in Table~\ref{Tab:Bckg}. 

The levels of cosmogenic backgrounds ($\beta$ + neutron decays of $^9$Li and $^8$He, fast neutrons, untagged muons) 
and of the background due to spontaneous fission in PMTs, did not change with respect to our previous paper~\cite{BorexGeonu}. 
We underline, that in order to suppress cosmogenic background we still apply a 2~s veto after a muon passes through the scintillator (mostly for $^9$Li and $^8$He) and 2~ms veto after muons pass through only the water tank (mostly for fast neutrons). These vetos induce about an 11\% loss of live time. In addition when possible, the pulse shape of the candidates was checked by an independent 400\,MHz digitizer acquisition system in order to further suppress undetected muon background.

The Borexino scintillator radioactivity has changed in time mostly because of the six purification campaigns performed in 2010 and 2011. During periods of no operations, $^{210}$Po, the main contaminant important for $\bar{\nu}_e$'s studies, is observed to decay exponentially with a $\tau$ = 199.6~days. The mean $^{210}$Po activity during the period used for this work is 15.8~counts/day/ton. Backgrounds from accidental coincidences and from ($\alpha$, $n$) interactions were evaluated according to the same methods as described in~\cite{BorexGeonu}.

During the purification campaigns some radon did enter the detector. The $^{222}$Rn has $\tau$ = 5.52 days and within several days the correlated backgrounds disappear. leaving in the detector the corresponding amount of $^{210}$Pb.
 These transition periods are not used for solar--$\nu$ studies, but, with special care can be used for $\bar{\nu}_e$ studies. 
 The $^{214}$Bi($\beta$) - $^{214}$Po($\alpha$) delayed coincidence has a time constant very close to the neutron capture time in PC. The $\alpha$ particles emitted by the $^{214}$Po usually show a visible energy well below the neutron capture energy window. However, in 1.04 $\times$ $10^{-4}$ or in 6 $\times$ $10^{-7}$ of cases, the $^{214}$Po decays to excited states of $^{210}$Pb and the $\alpha$ is accompanied by the emission of prompt gammas of 799.7~keV and of 1097.7~keV, respectively. In liquid scintillators, the $\gamma$ of the same energy produces more light with respect to an $\alpha$ particle~\cite{CTF}. Therefore, for these ($\alpha$ + $\gamma$) decay branches the observed light yield is higher with respect to pure $\alpha$ decays and is very close to the neutron capture energy window. 
 We have observed such candidates restricted to the purification periods, having the corresponding increased
   $Q_{\rm delayed}$ and positive ($\alpha$--like) Gatti parameter. In order to suppress this background to negligible levels during the purification periods, we have increased (with respect to~\cite{BorexGeonu}) the lower limit on $Q_{\rm delayed}$ to 860~p.e. and applied a slight Gatti cut on the delayed candidate as described above. 

\begin{table}[h]
\caption{Summary of the background faking anti--neutrino interactions and expressed in number of events expected among the 46 golden anti--neutrino candidates. The upper limits are given for 90\% C.L.}
\begin{center}
\begin{tabular}{lc}
\hline\hline
Background source			        	&Events \\
\hline
$^9$Li--$^8$He				&0.25$\pm$0.18 \\ 
Fast $n$'s ($\mu$'s in WT)		&$<$0.07 \\
Fast $n$'s ($\mu$'s in rock)		&$<$0.28 \\
Untagged muons				&0.080$\pm$0.007 \\
Accidental coincidences			&0.206$\pm$0.004 \\
Time corr. background			&0.005$\pm$0.012 \\
($\gamma$,n)					&$<$0.04  \\
Spontaneous fission in PMTs		&0.022$\pm$0.002 \\
($\alpha$,n) in scintillator	       &0.13$\pm$0.01 \\
($\alpha$,n) in the buffer		&$<$0.43\\
\hline
Total							&0.70 $\pm$ 0.18  \\ 
\hline\hline
\end{tabular}
\label{Tab:Bckg} 
\end{center}
\end{table}


We have identified 46 golden anti--neutrino candidates passing all the selection criteria described above, having uniform spatial and time distributions.
All prompt events of these golden candidates have a negative $G$ parameter, confirming that they are not due to $\alpha$'s or fast
protons. 
The total number of the expected background is (0.70 $\pm$ 0.18) events (see Table~\ref{Tab:Bckg}).
The achieved signal--to--background ratio of $\sim$65 is high due to the extreme radio--purity of Borexino scintillator and high efficiency of the detector shielding.

In the energy region $Q_{\rm prompt}$ $>$ 1300 p.e., above the end--point of the geo--neutrino spectrum, we observe 21 candidates, while the expected background as in Table~\ref{Tab:Bckg} is (0.24 $\pm$ 0.13) events. In this energy window, we expect (39.9 $\pm$ 2.7) and (22.0 $\pm$ 1.6) reactor--$\bar{\nu}_e$ events without and with oscillations, respectively. The expected survival probability is therefore (55.1 $\pm$ 5.5)\%, a value almost constant for distances $L_r$ $>$ 300~km.
We recall that for Borexino the closest reactor is at 416\,km and the mean weighted distance is 1200\,km.
We conclude that our measurement of reactor $\bar{\nu}_e$'s in terms of number of events is statistically in agreement with the expected signal in the presence of neutrino oscillations.
The ratio of the measured number of events due to reactor $\bar{\nu}_e$'s with respect to the expected non-oscillated number of events is (52.0 $\pm$ 12.0)\%.

We have performed an unbinned maximal likelihood fit of the light yield spectrum of our prompt candidates. The weights of the geo--neutrino (Th/U mass ratio fixed to the the chondritic value of 3.9~\cite{chondritic}) and the reactor anti--neutrino spectral components were left as free fit parameters. The main background components were restricted within $\pm$1$\sigma$ around the expected value as in Table~\ref{Tab:Bckg}. For the accidental background we have used the measured spectral shape,  while for the ($\alpha$, n) background we have used a MC spectrum. For the $^9$Li and $^8$He background we have used a MC spectrum as well which is in agreement with the measured spectrum of 148 events satisfying our selection cuts as observed within a 2~s time interval after muons passing the scintillator. 

\begin{figure}[h]
\centerline{\includegraphics[width=0.5\textwidth]{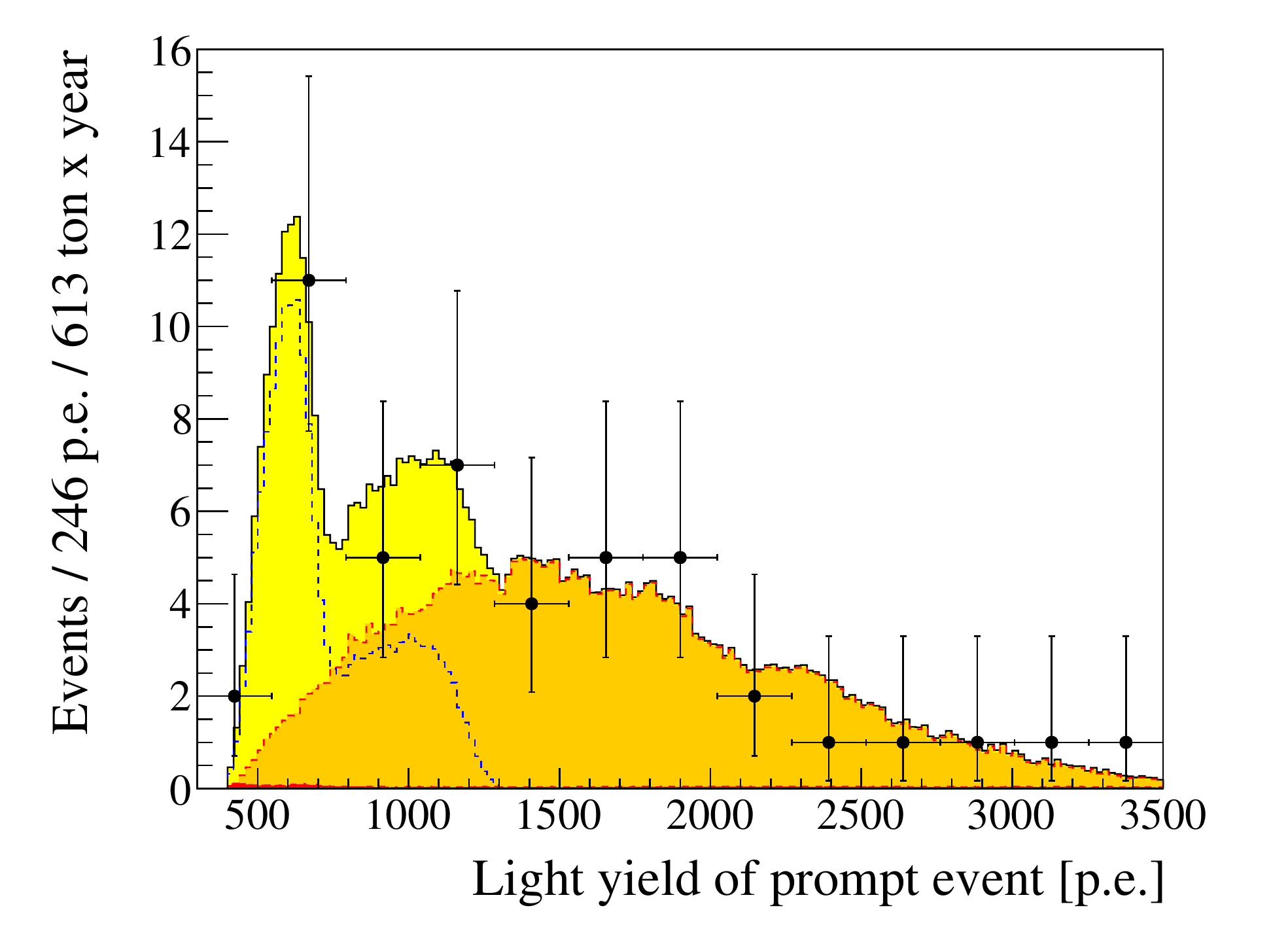}}
\caption{$Q_{\rm {prompt}}$ light yield spectrum of the 46 prompt golden anti--neutrino candidates and the best fit.  The yellow area isolates the contribution of the geo--$\bar{\nu}_e$ in the total signal.  Dashed red line/orange area: reactor--$\bar{\nu}_e$ signal from the fit. Dashed blue line: geo--$\bar{\nu}_e$ signal resulting from the fit. The contribution of background from Tab.~\ref{Tab:Bckg} is almost negligible and is shown by the small red filled area in the lower left part. The conversion from p.e. to energy is approximately 500 p.e./MeV. }
\label{Fig:fit}
\end{figure}

\begin{figure}[h]
\centerline{\includegraphics[width=0.45\textwidth]{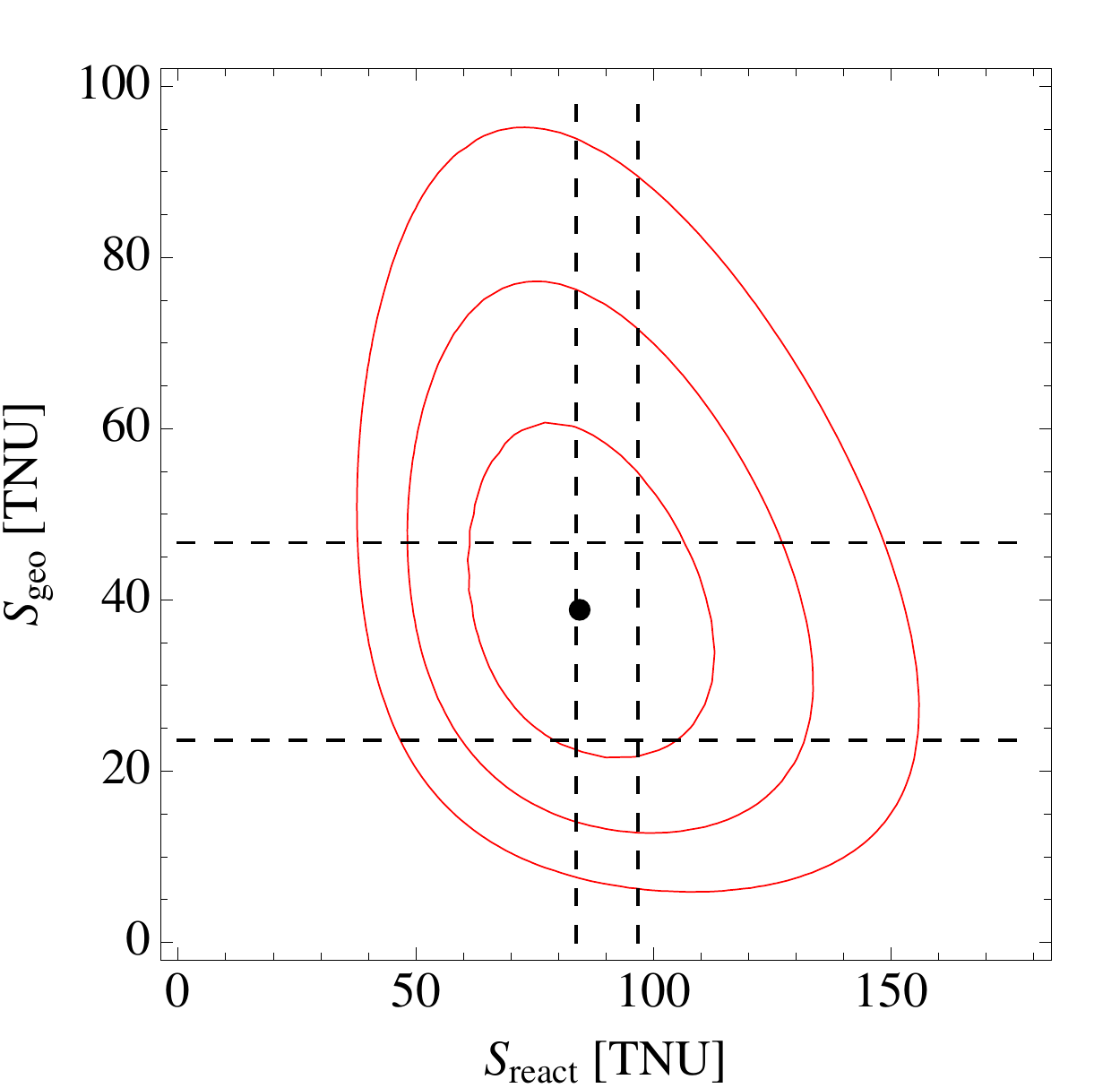}}
\caption{The 68.27, 95.45, and 99.73\% C.L. contour plots for the geo--neutrino and the reactor anti--neutrino signal rates expressed in TNU units. The black point indicates the best fit values. The dashed vertical lines are the 1$\sigma$ expectation band for $S_{rea}$. The horizontal dashed lines show the extremes of the expectations for different BSE models (see Fig.~\ref{Fig:BSE} and relative details in text).}
\label{Fig:contour}
\end{figure}

Our best fit values are $N_{\rm geo}$ = (14.3 $\pm$ 4.4) events and $N_{\rm react}$ = $31.2^{+7.0}_{-6.1}$ events, corresponding to signals $S_{\rm {geo}}$ = (38.8 $\pm$ 12.0)~TNU\footnote{1 TNU = 1 Terrestrial Neutrino Unit = 1 event / year / 10$^{32}$ protons} and $S_{\rm {react}}$ = 84.5$^{+19.3}_{-16.9}$~TNU. 
The measured geo--neutrino signal corresponds to overall $\bar{\nu}_e$ fluxes from U and Th decay chains of $\phi$(U) = $(2.4 \pm 0.7)  \times 10^6$ cm$^{-2}$ s$^{-1}$ and $\phi$(Th) = $(2.0 \pm 0.6)  \times 10^6$ cm$^{-2}$ s$^{-1}$, considering the cross section of the detection interaction (Eq.~\ref{Eq:InvBeta}) from~\cite{Strumia}.
From the $\ln{\cal{L}}$ profile, the null geo--neutrino measurement has a probability of 6 $\times$ $10^{-6}$.
The data and the best fit are shown in Fig.~\ref{Fig:fit}, while Fig.~\ref{Fig:contour} shows the 68.27, 95.45, and 99.73\% C.L. contours for the geo--neutrino and the reactor anti--neutrino signals in comparison to expectations. The signal from the reactors is in full agreement with the expectations of (33.3 $\pm$ 2.4) events in the presence of neutrino oscillations. 

A contribution of the local crust (LOC) to the total geo--neutrino signal, based on the local 3D geology around the LNGS laboratory, was carefully estimated in~\cite{Coltorti} as $S_{\rm geo}$(LOC) = (9.7 $\pm$ 1.3)~TNU.
The contribution from the Rest Of the Crust (ROC), based on the recent calculation by Huang et al.~\cite{Huang}, results in the geo--neutrino signal from the crust (LOC+ROC) of $S_{\rm geo}$(Crust) = (23.4 $\pm$ 2.8)~TNU. Subtracting the estimated crustal components from the Borexino geo--neutrino rate, we can infer the contribution of the mantle, $S_{\rm geo}$(Mantle) = (15.4 $\pm$ 12.3) TNU.

\begin{figure}[h]
\centerline{\includegraphics[width=0.5\textwidth]{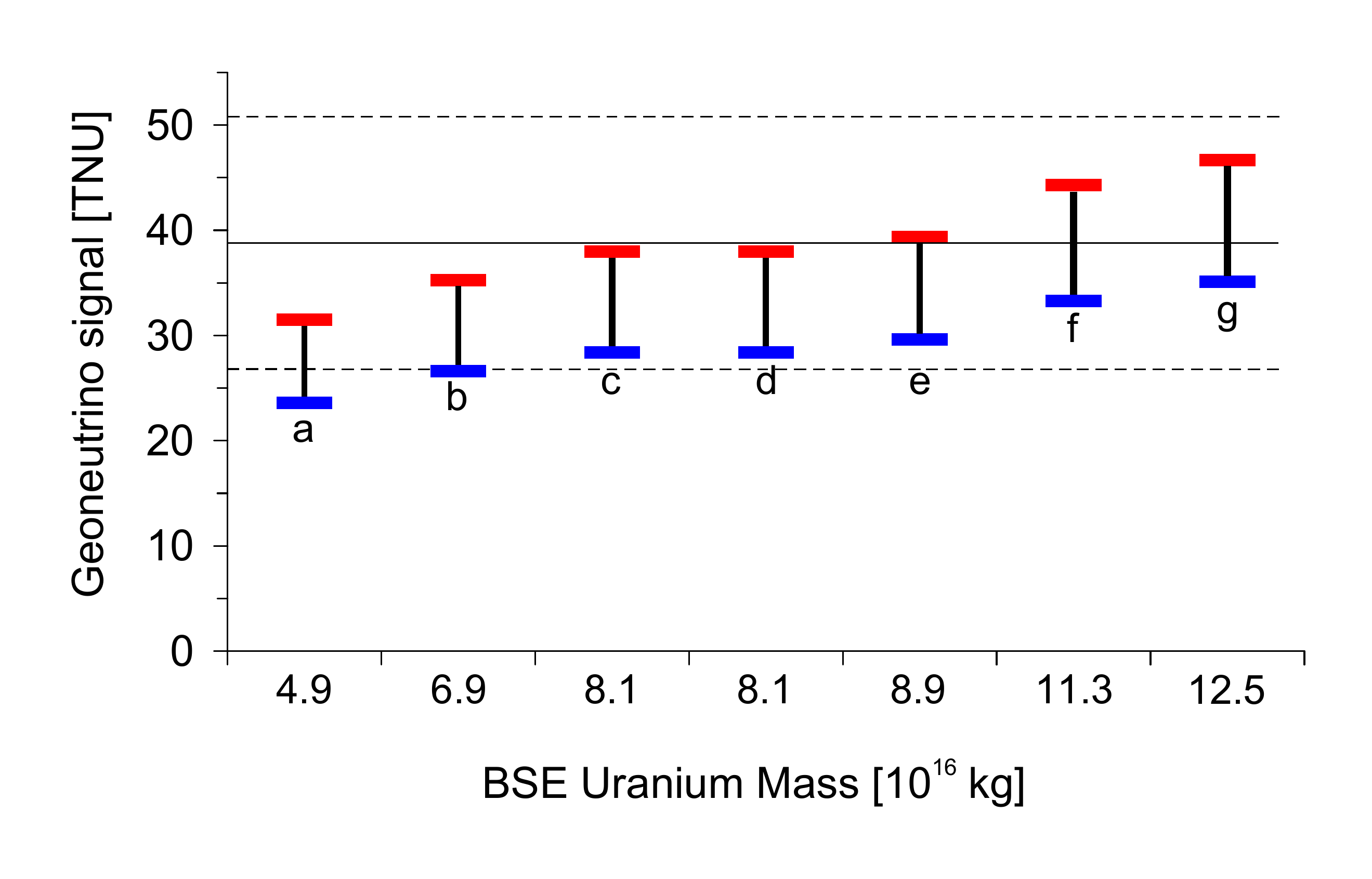}}
\vspace{-3 mm}
\caption{Geo--neutrino signal $S_{\rm geo}$ in Borexino (solid line) with $\pm$1$\sigma$ uncertainty (dashed lines) compared with the predicted values. The $\pm$1$\sigma$ band of $S_{\rm geo}$(LOC + ROC) crustal contribution~\cite{Fiorentini2012} is summed with $S_{\rm geo}$(mantle) according to seven BSE models: a) Javoy et al.~\cite{Javoy}, b) Lyubetskaya and Korenaga~\cite{Lyubetskaya}, c) McDonough and Sun~\cite{McDonough}, d) Allegre et al.~\cite{Allegre}, e) Palme and O’Neil~\cite{Palme}, f) Anderson~\cite{Anderson}, g) Turcotte and Schubert~\cite{Turcotte}. Red (blue) segments correspond to “high” (“low”) models obtained with two extreme distribution of U and Th in the mantle as described in the text, based on~\cite{Fiorentini2012}. On the $x$--axis we show the total uranium mass predicted by each BSE model in the primordial mantle. }
\label{Fig:BSE}
\end{figure}

On the basis of cosmochemical arguments and geochemical evidences, the different Bulk Silicate Earth (BSE) models predict the chemical composition of the Primitive Mantle of the Earth subsequent to the metallic core separation and prior to the crust--mantle differentiation. The predicted amount of U and Th in the mantle can be obtained by subtracting their relatively well known crustal abundances from the BSE estimates. The mantle geo--neutrino signal on the Earth surface depends not only on the absolute abundances of the radioactive elements but also on their distribution in the present mantle. For a fixed mass of U and Th, the extreme cases of $S_{\rm geo}$(Mantle) are obtained by distributing their abundances either homogeneously in the mantle (so called {\it high} model) or in an enriched layer close to the core--mantle boundary (so called “{\it low} model)~\cite{Fiorentini2007, Sramek}. In this perspective our results are summarized in Fig.~\ref{Fig:BSE}, which is obtained by combining the expected geo--neutrino signal from the crust (LOC + ROC) with those from different BSE models reported in Table~V of~\cite{Fiorentini2012}. The current result cannot discriminate among the different BSE models.

We have performed a combined analysis of our result with that of KamLAND~\cite{gando} in order to extract the $S_{\rm geo}$(Mantle).
First, the corresponding LOC + ROC crustal contributions taken from~\cite{Fiorentini2012} and \cite{Huang}, respectively, were subtracted from the measured $S_{\rm geo}$ signal: $S_{\rm geo}$(Crust) = (23.4 $\pm$ 2.8)~TNU for Borexino and $S_{\rm geo}$(Crust) = (25.0 $\pm$ 1.9)~TNU for KamLAND. 
Then, a spherically symmetric mantle was assumed.
The best fit value for the mantle signal common for both sites is $S_{\rm geo}$(Mantle) = (14.1 $\pm$ 8.1)~TNU.

\begin{figure}[h]
\centerline{\includegraphics[width=0.52\textwidth]{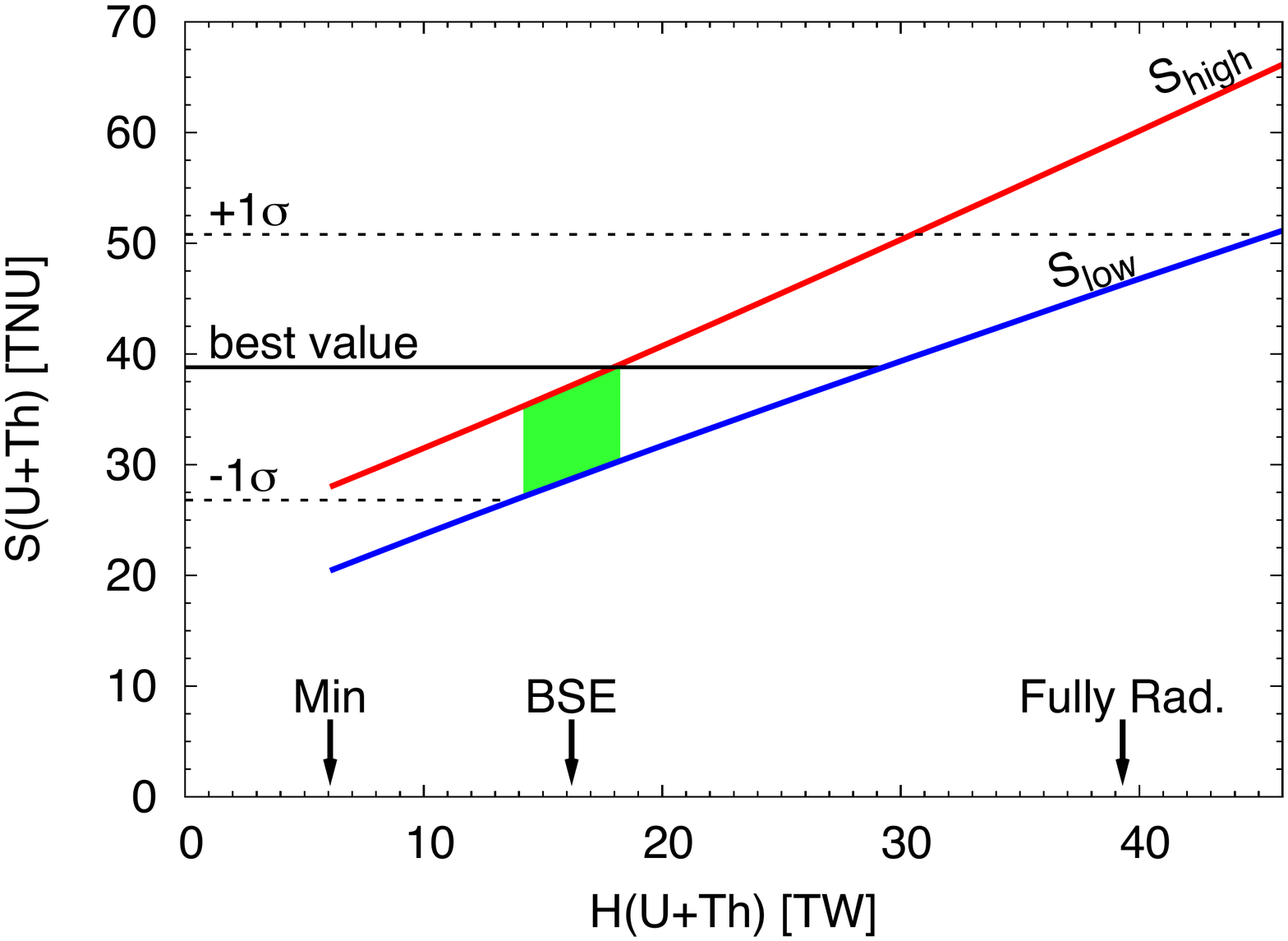}}
\vspace{-0.7 cm}
\centerline{\includegraphics[width = 0.52\textwidth]{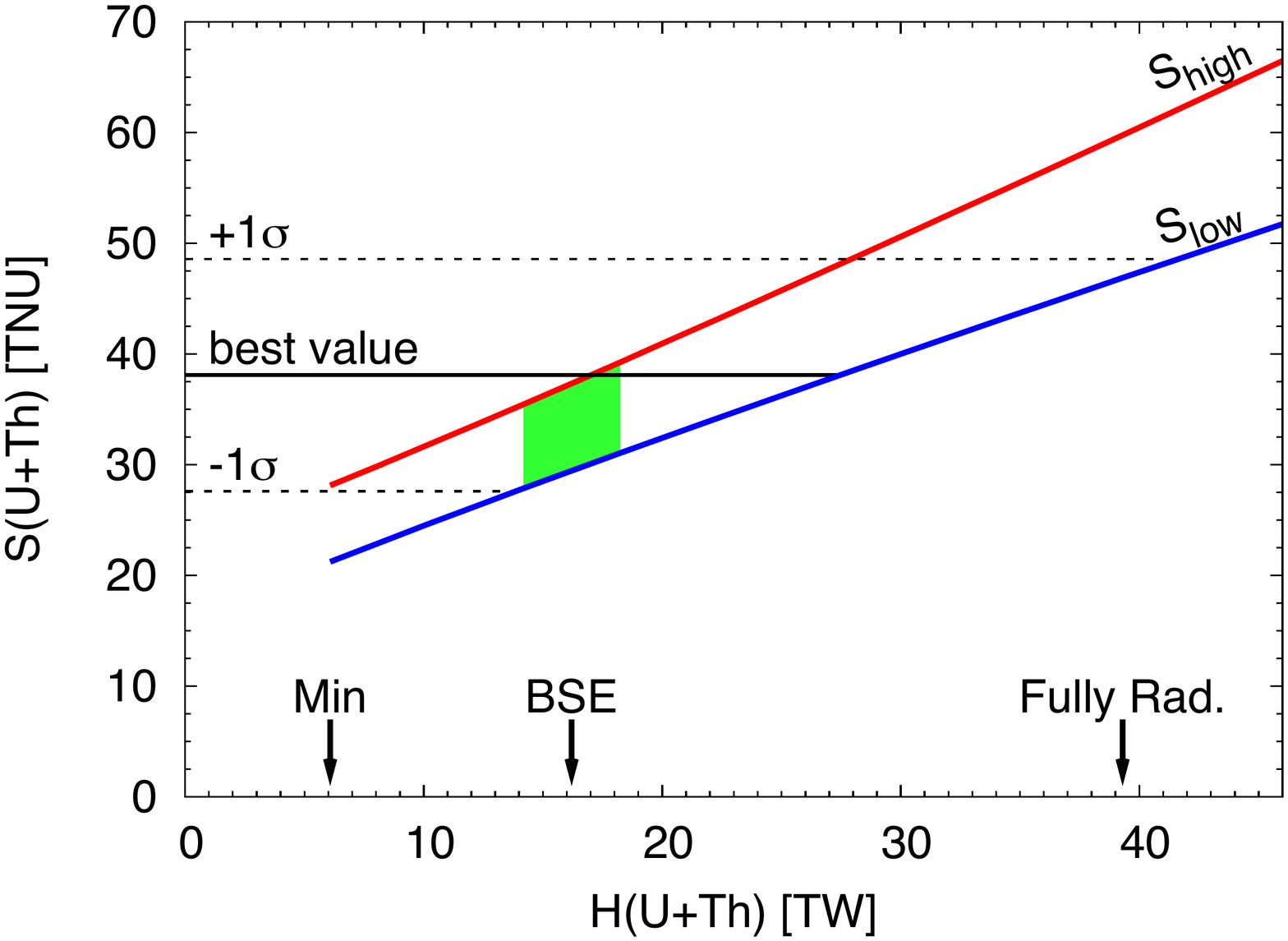}}
\vspace{-0.3 cm}
\caption{The signal $S_{\mathrm{U+Th}}$ from U and Th geo--neutrinos as a function of radiogenic heat production rate $H_{\mathrm{U+Th}}$ in Borexino (top) and KamLAND (bottom). Details in text.}
\label{Fig:signal_heat}
\end{figure}

The Earth releases radiogenic heat, $H_{\rm geo}$, together with geo--neutrinos in a well fixed ratio, however the observed geo--neutrino signal depends both on the abundances of the individual radioactive elements and on their distribution inside the Earth.
To extract the radiogenic heat power from a measured $S_{\rm geo}$ is therefore model dependent.
We have calculated the expected $S_{\rm geo}$(U+Th) as a function of the radiogenic heat produced by U and Th, $H_{\rm geo}$(U+Th), for the Borexino and KamLAND sites (see Fig.~\ref{Fig:signal_heat}), and compared it to the Borexino and KamLAND~\cite{gando} results.
The allowed regions between the red and blue lines in the plane $S_{\rm geo}$(U+Th) and $H_{\rm geo}$(U+Th) contain models consistent with geochemical and geophysical data.
For each total mass of U and fixed Th/U ratio, the maximal geo--neutrino signal (red line) can be obtained by maximizing the radiogenic material in the crust and allowing uniform distribution in the mantle. 
Similarly, the minimal signal (blue line) is obtained for the minimal radiogenic mass in the crust with the rest concentrated in a thin layer at the bottom of the mantle.
The expected signal from the crust is taken from Table V of~\cite{Fiorentini2012}.
We have chosen as a reference the BSE model from~\cite{McDonough}, predicting that the silicate Earth contains m(U) = (0.8 $\pm$ 0.1) $\times$ $10^{17}$~kg with mass ratios Th/U = 3.9 and K/U = 12000.
The green regions are allowed by the BSE model~\cite{McDonough}. 
The arrow "Min" indicates the contribution of the crust only.
The arrow for the fully radiogenic model indicates 39.3~TW: it assumes that the total Earth surface heat flux of (47 $\pm$ 2)~TW ~\cite{Davies} is completely due to radiogenic heat from U, Th, and K.
Taking the relative proportions from the BSE of~\cite{McDonough}, we get that in a fully radiogenic Earth, U, Th, and K produce 19.1, 20.2, and 7.7~TW, respectively. 

\begin{figure}[t!]
\centerline{\includegraphics[width=0.59\textwidth]{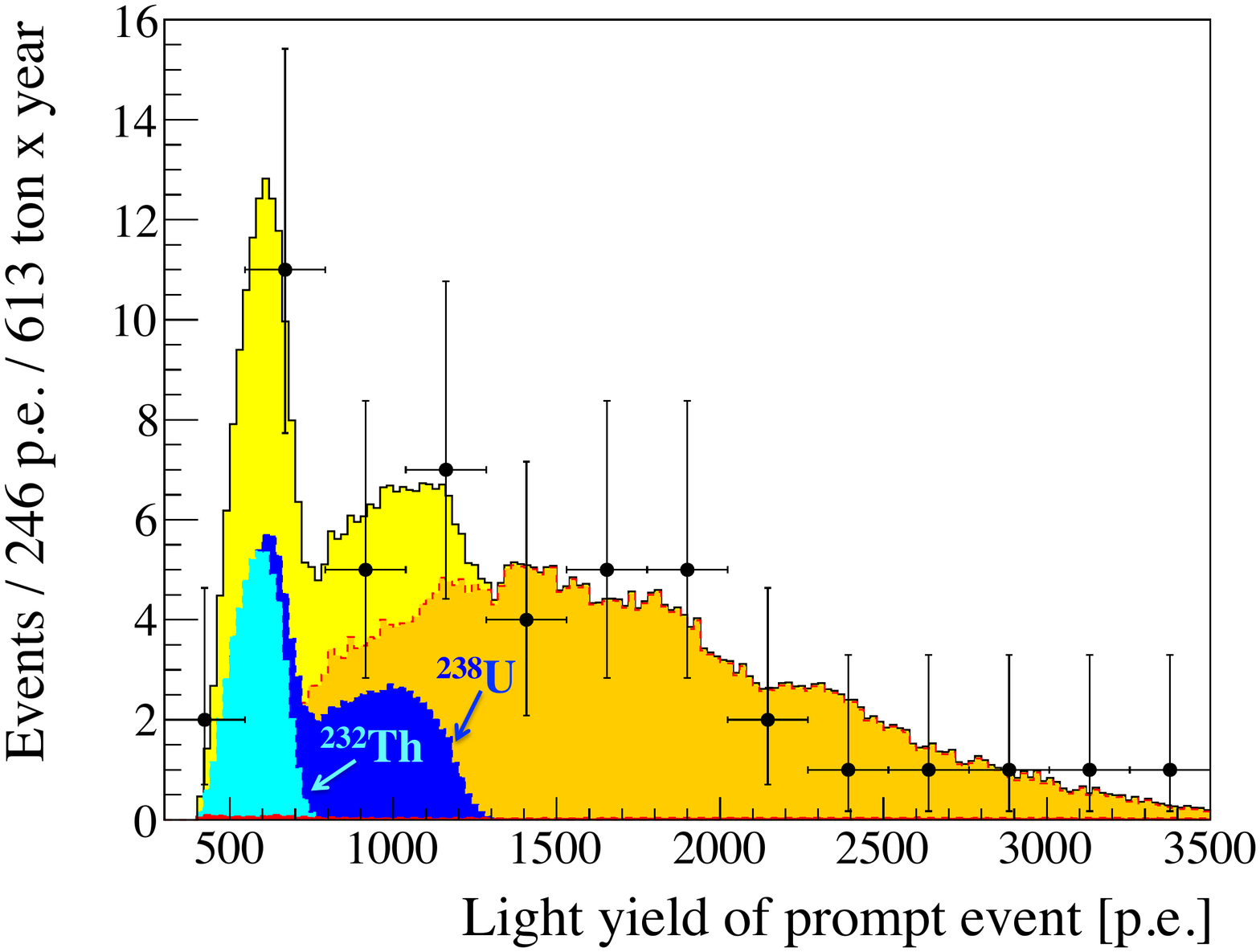}}
\vspace{-0.4 cm}
\caption{$Q_{\rm {prompt}}$ light yield spectrum of the 46 prompt golden anti--neutrino candidates and the best fit with free U (blue) and Th (cyan) contributions.  The yellow area isolates the total contribution of geo--$\bar{\nu}_e$s. Dashed red line/orange area: reactor--$\bar{\nu}_e$ signal from the fit. The contribution of background from Tab.~\ref{Tab:Bckg} is almost negligible and is shown by the small red filled area. The conversion from p.e. to energy is approximately 500 p.e./MeV. }
\label{Fig:fit_UThfree}
\end{figure}

\begin{figure}[t!]
\centerline{\includegraphics[width = 0.45\textwidth]{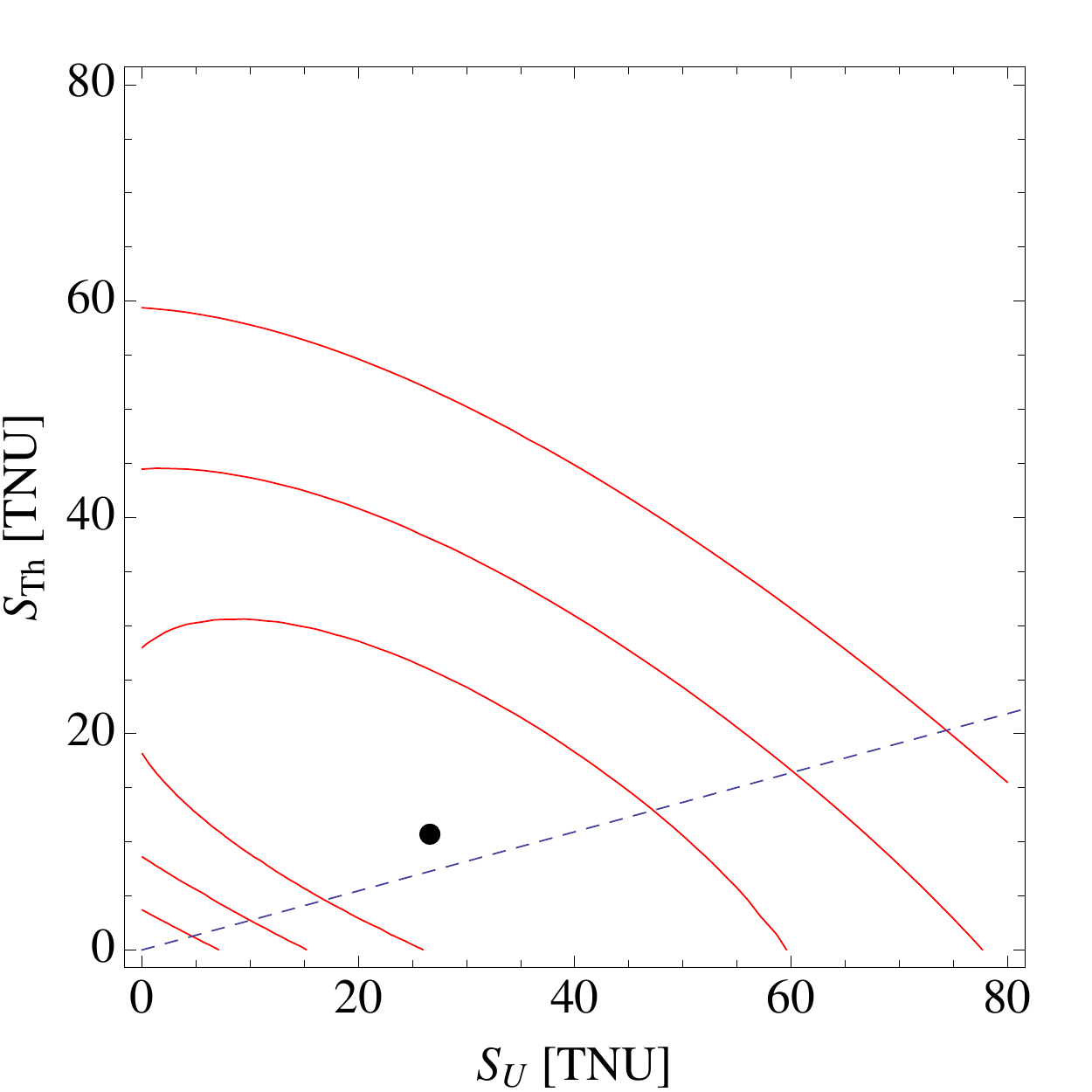}}
\caption{The 68.27, 95.45, and 99.73\% C.L. contour plots of the $S_{\mathrm{Th}}$ and $S_{\mathrm{U}}$ signal rates expressed in TNU units. The black point indicates the best fit values. The dashed blue line represents the chondritic Th and U ratio.}
\label{Fig:contourUTh}
\end{figure}

We have performed another unbinned maximal likelihood fit of our 46 golden candidates in which the individual contributions from the $^{238}$U and $^{232}$Th chains were fitted individually (see Fig.~\ref{Fig:fit_UThfree}), with all other fit details as above. 
The best fit values are $N_{\mathrm{Th}}$ = (3.9 $\pm$ 4.7)~events and $N_\mathrm{U}$ = (9.8 $\pm$ 7.2)~events, corresponding to $S_{\mathrm{Th}}$ = (10.6 $\pm$ 12.7)~TNU and $S_\mathrm{U}$ = (26.5 $\pm$ 19.5)~TNU and $\bar{\nu}_e$ fluxes (above 0\,MeV) of $\phi$(Th) = $(2.6 \pm 3.1)  \times 10^6$ cm$^{-2}$ s$^{-1}$ and $\phi$(U) = $(2.1 \pm 1.5)  \times 10^6$ cm$^{-2}$ s$^{-1}$.
The  68.27, 95.45, and 99.73\% C.L. contour plots of $S_{\mathrm{Th}}$ versus $S_{\mathrm{U}}$ are shown in Fig.~\ref{Fig:contourUTh}.
Although our data is compatible within 1$\sigma$ with only $^{238}$U signal (and $S_{\mathrm{Th}}$ = 0) or only $^{232}$Th signal (and $S_{\mathrm{U}}$ = 0), we note that the best fit of the Th/U ratio is in very good agreement with the chondritic value.

A geo--reactor with thermal power $<$30~TW and $^{235}$U : $^{238}$U = 0.76 : 0.23 composition was suggested by Herndon~\cite{Herndon2}.
It is assumed to be confined in the central part of the Earth's core within the radius of about 4~km~\cite{Herndon}.
We have produced MC spectra of the expected geo--reactor anti--neutrino. 
In a similar unbinned maximal likelihood fit of our 46 golden anti--neutrino candidates we have added another fit component, $N_{\rm geo-react}$, while constraining $N_{\rm react}$ to the expected value of (33.3 $\pm$ 2.4) events.
All other fit details were as above, including fixed chondritic  mass Th/U ratio.
We set the upper limit on the geo--reactor power 4.5~TW at 95\%~C.L.

We are extremely grateful to E. Padovani, for  information about CANDU fuel, to J. Mandula for providing us detailed data on monthly load factor for each nuclear cores in the world, and to G. Alimonti, M. Lissia, W. F. McDonough for  useful discussions.
The Borexino program is made possible by funding from INFN (Italy), NSF (USA), BMBF, DFG (OB 168/1-1), MPG, and the Garching accelerator laboratory MLL (Germany), Russian Foundation for Basic Research (Grant 12-02-12116) (Russia), MNiSW (Poland), and the UnivEarthS LabEx programme (ANR-11-IDEX-0005-02) (France). We acknowledge the generous support of the Gran Sasso National Laboratory (LNGS).


\end{document}